\documentclass[11pt]{article}
\usepackage{graphicx,floatflt,amssymb,epsf,psfig,rotate} 
\def\DESepsf(#1 width #2){\epsfxsize=#2 \epsfbox{#1}}

\def\be{\begin{equation}}
\def\ee{\end{equation}}
\def\ba{\begin{eqnarray}}
\def\ea{\end{eqnarray}}

\def\br{\begin{array}}
\def\er{\end{array}}
\def\bc{\begin{center}}
\def\ec{\end{center}}

\def\std{{\cal G}_{std}}
\begin{document}
\thispagestyle{empty}
\begin{flushright} 
HRI-P-08-07-002 \\
CU-Physics/11 - 2008\\ 
\end{flushright}  
\vskip 10pt 
\begin{center}
{\Large \bf Neutrino mass and low-scale leptogenesis
in a testable SUSY SO(10) model\\} \vskip .1in

\bf{\sf Swarup Kumar Majee${}^{1}$, Mina K. Parida${}^2$, Amitava
Raychaudhuri${}^{1,3}$}
\vskip .20in
{\sl ${}^1$ Harish-Chandra Research Institute, Jhunsi, Allahabad
211 019, India}\\
{\sl ${}^2$ National Institute of Science Education and Research,
Institute of Physics Campus,\\ Sachivalaya Marg, Bhubaneswar 751
005,  India}\\
{\sl ${}^3$ Department of Physics, University of Calcutta,
Kolkata 700 009, India}

\end{center}
\vskip .25in

\begin{abstract}
\noindent
   It is shown that a  supersymmetric  ~$\bf {SO(10)}$  model
extended with fermion singlets can accommodate the observed
neutrino masses and mixings as well as generate the desired
lepton asymmetry in concordance  with the gravitino constraint.
A necessary prediction of the model is near-TeV scale
doubly-charged Higgs scalars which should be detectable at the
LHC.


\end{abstract}


The observed neutrino masses constitute a compelling evidence of
interactions beyond 
 the Standard Model of particle physics and leave
an impact in areas as diverse as astrophysics, cosmology,
nuclear physics, and geophysics. The smallness of these masses
finds a natural explanation in the see-saw mechanism
\cite{seesaw}, which requires a heavy Majorana (self-conjugate)
neutrino. Such heavy neutrinos appear in grand unified theories
(GUTs) based on $\bf {SO(10)}$, which incorporate 
quark-lepton unification  and left-right symmetry \cite{ps,lr}.  
The wide disparity between the weak and unification
scales in these models calls for a protection mechanism and
supersymmetry (SUSY) is widely considered to be an attractive
candidate. Further, such a model with a low SUSY scale leads to a
unification of gauge couplings at high energies.  These positive
features have encouraged many explorations of the SUSY  $\bf
{SO(10)}$ model.

Another open problem, also of much interest, is the origin of the
observed baryon asymmetry of the universe. Originally it was
expected that baryon number violation inherent in GUTs will lead
to this small asymmetry when heavy gauge (and/or Higgs) bosons
decay while they are out of equilibrium.  This hope was belied
however since any primordial GUT-origin asymmetry will be totally
diluted in the inflationary epoch. This has provided impetus to
look for lower energy avenues for generating this asymmetry.
An oft-chosen route is to generate a lepton asymmetry through the
C and CP-violating out-of-equilibrium decay of heavy Majorana
neutrinos. This is later converted to a baryon asymmetry through
anomalous $(B+L)$ violation, which is implicit in the Standard
Model \cite{yana, lep2bar}.

It is but natural to ask whether the heavy Majorana neutrino
which drives the neutrino mass see-saw can also generate the
lepton asymmetry through its decay. This would have been truly
economical.

Hindrances to this programme arise from several directions. (a) The
observed light neutrino masses require the heavy neutrino, which
is right-handed (RH), to have a 
mass $\sim 10^{13}$ GeV. This sets the
scale, $M_R$, for the $\bf {SU(2)_R \times SU(2)_L \times
U(1)_{(B-L)} \rightarrow SU(2)_L}$ $\bf {\times U(1)_Y}$ gauge
symmetry breaking.  (b) Within the $\bf {SO(10)}$ GUT
framework, the intermediate symmetry breaking scales are fixed through
the Renormalization Group (RG) equations which reflect the gauge
couplings' evolution with energy. In the simplest $\bf {SO(10)}$
GUT it is well-known that $M_R$ turns out to be $\sim 10^{16}$
GeV. (c) In a SUSY context there is an additional constraint,
namely, to ensure that there is no overabundance of gravitinos in
the universe. To maintain consistency with this, it has been
demonstrated \cite{lepto, khlopov, di} that the lepton asymmetry
must be generated through the decay of a heavy neutrino whose
mass does not exceed  $\sim 10^{7-9}$ GeV in order to prevent a
washout, whereas leptogenesis through the canonical Type-I
see-saw mechanism sets the lower bound $4.5 \times 10^9$ GeV.
These conflicting requirements have acted as obstacles to a
successful implementation of this attractive possibility.

In this letter we propose a remedy for these maladies confining
ourselves to the SUSY $\bf {SO(10)}$ GUT.  If sterile -- i.e.,
$\bf {SO(10)}$ singlet -- leptons are introduced, one for each
generation \cite{e6, vm, barr, kk}, then a novel way can be found to
meet the demands outlined in the previous paragraph.

The uncharged fermions in this model, per generation,  are the
following: a left-handed neutrino $\nu$, a right-handed neutrino,
$N$, and a sterile neutrino, $S$.  For the three generation
neutral fermion system, the mass matrix on which we focus is:
\ba
M_\nu = \pmatrix{\nu & N^c & S}_L \pmatrix{ 0 & m_D & 0  \cr
m_D^T & M_N & M_X\cr 0 & M_X^T & \mu  }
\pmatrix{ \nu^c \cr N \cr S}_R 
\label{matrix}
\ea
where $m_D, M_N, M_X$, and  $\mu$ are all 3$\times$3 matrix blocks.

It is not unreasonable to expect that the mass matrix in eq.
(\ref{matrix}) will alleviate the tension,
summarised earlier, between light neutrino masses and adequate
low-scale thermal leptogenesis. As discussed below, the double
see-saw structure for the  light neutrino masses,
arising from eq. (\ref{matrix}), also decouples it to some extent from
low-scale leptogenesis;  $M_N, M_X$, and $\mu$ appear in
different fashions in the expressions.  Utilizing an extension of
the Minimal Supersymmetric Standard Model (MSSM) by the
addition of RH neutrinos and extra fermion singlets, which results in
a neutrino mass matrix of the structure of eq.  (\ref{matrix}), Kang and
Kim \cite{kk} have found solutions to both the above issues.
There,  $m_D$ has been identified, as is done in the MSSM, with
the charged lepton mass matrix.  On the other hand,  in the $\bf
{SO(10)}$ model which is espoused here, quark-lepton symmetry
\cite{ps} identifies  the neutrino Dirac mass matrix $m_D$ with
the up-quark mass matrix whose 33 element is nearly 100 times
heavier. This, along with other GUT constraints, pose additional
hurdles in addressing the problems in SUSY $\bf {SO(10)}$.

 We work in a basis in which the down-quark and charged lepton
mass matrices are diagonal. This ensures that the entire mixings
in the quark and lepton sectors can be ascribed to the mass
matrices of the up-type quarks and the neutrinos, respectively.
Using quark-lepton unification, the quark masses, and the
Cabibbo-Kobayashi-Maskawa mixing angles, one therefore obtains
$m_D$, upto $\cal O$(1) effects due to RG evolution.

We utilize spontaneous symmetry breaking of SUSY  $\bf {SO(10)}$
with the Higgs representations ${\bf {210}, {54}}$, ${\bf 126
\oplus \overline {126}, 16\oplus \overline {16}}$, and $\bf
{10}$. By using the mechanism of D-Parity breaking near the GUT
scale \cite{dpar}, the RH-triplet pair  $\Delta_R(1,3,1,-2)
\oplus \overline{\Delta}_R(1,3,1,2)$, and the RH-doublet pair
$\chi_R(1,2,1,-1) \oplus \overline{\chi}_R(1,2,1,1)$ are treated
to have masses at much lower scales compared to their left-handed
counterparts.  $M_X = F x_R$, in eq. (\ref{matrix}), is generated
{\em via} the vacuum expectation value (vev)
$<\chi_R(1,2,1,-1)> =$ $<\overline{\chi}_R(1,2,1,1)> = x_R$,   
where we take $F$ to be a matrix with entries $\cal O$(0.1).

 Although we do not assign any direct vev to the RH-triplets,
through a $\Delta_R$ exchange involving a trilinear coupling in
the superpotential, $\lambda \Delta_R \overline{\chi}_R
\overline{\chi}_R$, an effective vev $
{<\Delta_R(1,3,1,-2)> \equiv v_R = \lambda
\frac{x_R^2}{m_{\Delta_R}}}$ is generated, resulting in the mass
term $M_N \sim f <\Delta_R(1,3,1,-2)>$, where $f$ is a typical
Yukawa coupling of Majorana type. If ${m_{\Delta_R}}$ is around 1
TeV, which can be arranged by a  tuning of the D-parity breaking
term in the Lagrangian, the entries of $M_N$ are ${\cal
O}(10^{11})$ GeV.  Without any loss of generality, $M_N$ can be
chosen to be diagonal.

Notice that the {\bf $SU(2)_R\times U(1)_{B-L}$} symmetry breaks at the scale
\mbox{$<\Delta_R(1,3,1,-2)>$} $\simeq 10^{11}$ GeV while $x_R
\sim 10^7$ GeV. The states $\Delta^+_R$ and $Re(\Delta^0_R)$ are
eaten up as Goldstone bosons by the $W_R^\pm$ and  $W_R^0$ fields and
$\Delta^{++}_R$ and $Im(\Delta^0_R)$ survive as physical states
with mass $\sim$ 1 TeV. 
 The Type-II seesaw contribution to the light neutrino mass matrix is 
damped out in this case because of the large masses of the left-handed
Higgs triplet  leading to $m_{II} \simeq 10^{-5}$ eV -- $ 10^{-6}$
eV \cite{typeII,cm}. 
Further, the vev of $\chi_L$ is zero or negligible.

Block diagonalization of
the mass matrix in eq. (\ref{matrix}) in the limit in which we
are working ({\em i.e.,} $M_N \gg M_X \gg \mu \gg m_D$) leads to:
\begin{equation}
m_{\nu} \sim - {m_D}\,\left[{M_X}^{-1}\mu
({M_X}^T)^{-1}\right]\,{m_D}^T, \,\,\,\, M_S \sim \mu -
\frac{M_X^2}{M_N},
 \,\,\,\,M \sim M_N + \frac{M_X^2}{M_N},
\label{double}
\end{equation}
where $m_\nu, M_S$, and $M$ are $3\times3$ matrices. The light
neutrino masses are in a double see-saw pattern and $\mu$ is
determined once $M_X$ is fixed. It may be noted that the mass
matrix structure in eq. (\ref{matrix}) ensures that the type I
see-saw contribution is absent  and $M_N$ remains unconstrained
by the light neutrino masses. This freedom in $M_N$ -- a hallmark
of the model -- is vital to ensure adequate leptogenesis.
\begin{figure}[tbh]
\hskip 2.5cm
\psfig{figure=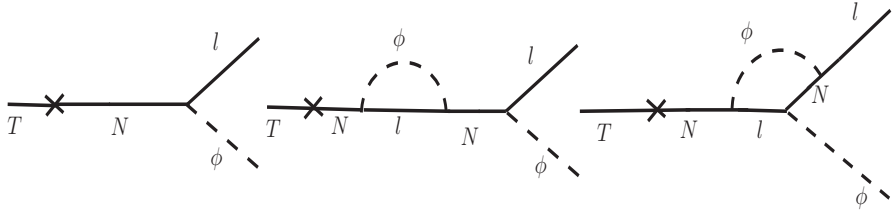,width=12.0cm,height=3cm,angle=0} 
\caption{\sf \small The tree and one-loop  contributions to the
decay of $T_1$ that generate the lepton asymmetry.}
\label{f1}
\end{figure}

The eigenstates of $M_S$, which we denote by $T_i, (i=1,2,3)$,
are superpositions of the sterile neutrinos $S$ (predominant) and
the right-handed ones $N$. These states are found to lie
well-below $10^9$ GeV, consistent with the gravitino constraint.
In fact we show that this model allows successful leptogenesis at
a temperature ${\rm T}\simeq M_T \simeq 5\times 10^{5}$ GeV, which is
nearly 4 orders below the maximum allowed value.  Further, the
singlet fermions  decay through their mixing  with the $N_i$
which is  controlled by the ratio $M_X/M_N$. The latter, which
have masses ${\cal O}(10^{11})$ GeV and are off-shell, decay to a
final $l \phi$ state, where $l$ is a lepton doublet and $\phi$
the up-type MSSM Higgs doublet.  This two-step process -- for
which  tree and loop diagrams  are depicted in Fig. \ref{f1}
(SUSY contributions are small) --
results in a lepton asymmetry of the correct order. Because of
the large value of $M_N \gg M_X$, a small $S-N$ mixing results
naturally  in the $T_i$ which in turn guarantees the
out-of-equilibrium condition to be realized near temperatures
${\rm T}\simeq M_T$.

A quantitative analysis of this programme has been carried out
using the Boltzmann equations determining the number densities
in a co-moving volume $Y_{T} = n_{T}/n_S$ and $Y_L =
n_L/n_S$, where $n_T$, $n_L$ and $n_S$ are respectively the number
densities of the decaying neutrinos, leptons and the entropy:
\be
\frac{dY_{T}}{dz}=-\left(Y_{T}-Y_T^{eq}\right) \left[ \frac{\Gamma_D^T}
{zH(z)}+\frac{\Gamma_s^T}{zH(z)} \right],\,\,
\frac{dY_{L}}{dz}=\epsilon_T \frac{\Gamma_D^T}{zH(z)} \left( Y_{T}-Y_T^{eq}
\right)-\frac{\Gamma_W^\ell}{zH(z)}Y_L.
\label{BE}
\ee
where $\Gamma_D^T$, $\Gamma_s^T$ and $\Gamma_W^\ell$ represent
the decay, scattering, and wash-out rates, respectively, that
take part in establishing a net lepton asymmetry. We refrain from
presenting their detailed expressions here \cite{mpr}. The Hubble
expansion rate $H(z)$, where $z = M_T/{\rm T}$, and the CP-violation
parameter are given by
\be
H(z)=\frac{H(M_T)}{z^2}, \,\,\,\, H(M_T)=1.67 g_*^{1/2}
\frac{M_T^2}{M_{pl}},\;\;\; \epsilon_T = \frac{\Gamma (T\rightarrow l
\phi) - \Gamma (T\rightarrow \bar{l}
\phi^*)}{\Gamma (T\rightarrow l
\phi) + \Gamma (T\rightarrow \bar{l}
\phi^*)}.
\label{hubble_eps}
\ee
Our target is to use eqs. (\ref{double}) and (\ref{BE}) to
obtain an acceptable solution within the framework of SUSY $\bf
{SO(10)}$.  Through an exhaustive analysis we find an appropriate
choice of the block matrices appearing in eq. (\ref{matrix})
which guarantees adequate leptogenesis while maintaining full
consistency with the observed neutrino masses and mixing as well
as the gravitino constraint. The mass scales are fixed as
dictated by the RG evolution of gauge couplings in SUSY $\bf {SO(10)}$
 when effects of two $\rm {dim}.5$ operators  scaled by the Planck mass
are included \cite{mpr,mprs,d5}. The strategy we follow is to choose 
the matrix $M_X$ first.
To minimize the number of independent parameters, we take the
matrix $F$ to be real and diagonal, which is reflected in $M_X$.
Then using $m_D$, as fixed by quark-lepton unification, $\mu$ is
determined from the double see-saw formula given in  eq. (\ref{double}).
Using these inputs, one has to examine, by trial and error,
different choices of $M_N$ for adequate lepton asymmetry
generation.
\begin{figure}[bth]
\hskip 3.5cm
\psfig{figure=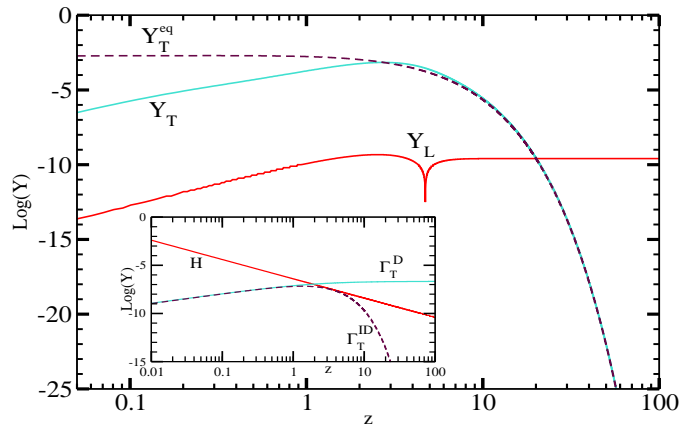,width=10.0cm,height=7.0cm,angle=270} 
\caption{\sf \small
The comoving density of $T_1$ -- $Y_T$ --  and the leptonic
asymmetry -- $Y_L$ -- as a function of $z$. Also shown is
$Y_T^{eq}$.  The inset displays the decay ($\Gamma_T^D$) and
inverse-decay ($\Gamma_T^{ID}$) rates of $T_1$ compared with the
Hubble expansion rate, $H$, as a function of $z$. }
\label{f2}
\end{figure}

The results for the development of the leptonic asymmetry as the
universe evolves 
 are
shown in Fig. \ref{f2}.  They are obtained with the choice $M_X$
=  diag (0.2, 0.3, 0.4)$ x_R$ with $x_R = 6 \times 10^6$ GeV. The
neutrino Dirac mass matrix, $m_D$, is constructed utilizing
quark-lepton symmetry; the up-type quark mass eigenvalues and the
CKM mixings are taken at the PDG \cite{pdg} values with the
CKM-phase  as 1 radian. The neutrino masses are fixed so as to
satisfy $\Delta m^2_{21} = 8.0
\times 10^{-5}$ eV$^2$ and  $\Delta m^2_{32} = 2.5 \times
10^{-3}$ eV$^2$ with the lightest neutrino taken massless.  The
neutrino mixing angles used are $\theta_{23} = 45^\circ$,
$\theta_{12} = 32^\circ$, and $\theta_{13} = 7^\circ$. No other
CP-phases are introduced in the lepton sector except the one
through the CKM matrix for quarks. With these inputs the matrix
$\mu$ is calculated following eq. (\ref{double}). For the
RH-neutrino we use the mass matrix  $M_N = {\rm diag}(0.1, 0.5,
0.9) \times 10^{11}$ GeV. This is consistent with $m_{\Delta_R}
\sim$ 1 TeV. We assume that in the very initial stages the number
densities, $Y_{T_i}, \,i=1,2,3$, and the leptonic asymmetry, $Y_L$, are zero.
The chosen input values of the mass parameters result in a $T_i$
mass spectrum such that only one state -- $T_1$ -- is above the
kinematic threshold for $l \phi$ production  ($m_{T_1} = 3.9
\times 10^5$ GeV) and the lepton asymmetry results through its
decay.  This ensures that the leptogenesis is consistent with the
gravitino bound.  It is seen from Fig. \ref{f2}  that $T_1$ decays
fall out of equilibrium as the universe expands (inset) and $Y_L$
achieves the right order ($\sim 10^{-10}$) starting off from a
vanishing initial value while $Y_{T}$ steadily tends towards
$Y^{eq}_{T}$. 

We stress again that an important outcome of the symmetry
breaking is that  out of the triplet $\Delta_R$ the
components $\Delta_R^\pm$ and $Re
\Delta_R^\circ$ are absorbed as longitudinal modes of the broken
generators of $SU(2)_R \times U(1)_{B-L}$. The physical states
are $\Delta_R^{++}$ and $Im
\Delta_R^\circ$ and their superpartners. They will be within
striking range of the LHC and the ILC with $m_{\Delta} \simeq
300$ GeV --  1 TeV.

Finally, we briefly discuss the mechanism of SUSY  $\bf {SO(10)}$ breaking
\cite{mpr,mprs,d5}: 
\begin{eqnarray}
 SO(10) & \stackrel{\mathbf{M_U}}{\longrightarrow} &
SU(3)_C \times SU(2)_R \times SU(2)_L \times U(1)_{B-L}
~~[{\cal G}_{3221}] \nonumber \\
&\stackrel{\mathbf{M_R}}{\longrightarrow}& SU(3)_C \times
SU(2)_L \times U(1)_Y ~~[\std]
\stackrel{\mathbf{M_Z}}{\longrightarrow}SU(3)_C \times U(1)_Q 
\label{chain}
\end{eqnarray}
The $\bf {SO(10)}$ Higgs multiplets {\bf 210} and {\bf 54} are
utilized to break the symmetry at $M_U$. Within the {\bf 210}
there are two components which develop vevs; one breaks $\bf
{SO(10)}$ to ${\cal G}_{3221}$ while the other is responsible for
D-parity breaking. The vev of the singlet under the Pati-Salam
group contained in {\bf 54} ensures that there are no light
pseudo-goldstone bosons arising from the {\bf 210} to upset
perturbative gauge coupling evolution. As already discussed, $\bf
{SU(2)_R}\times U(1)_{B-L}$ is broken by the vevs of RH-triplets
in {\bf 126} $\oplus$  {\bf $\overline{\bf 126}$}.  This  induced
vev, $v_R \sim 10^{11}$ GeV,  is also responsible for the masses
of the $N_i$.  The last step of breaking in eq. (\ref{chain})
relies on the electroweak vev of the weak bi-doublet  in {\bf
10}. We have carried out an analysis of the RG evolution of the
gauge couplings to determine the intermediate mass scales. We
find that  $M_R \sim 10^{9-11}$ GeV can be obtained through the
introduction of effective $\rm {dim}.5$ operators scaled by the
Planck mass, $M_{Pl}$ \cite{d5}.  It is noteworthy that both {\bf
210} and {\bf 54} are necessary for a viable SUSY $\bf {SO(10)}$
breaking pattern and that the resulting two $\rm {dim}.5$
operators are instrumental in alleviating the problem of
leptogenesis under the gravitino constraint:
\ba
{\cal L}_{NRO} &=& -{\eta_1\over
2M_{Pl}}Tr\left(F_{\mu\nu}\Phi_{210}F^{\mu\nu}\right) -{\eta_2\over
2M_{Pl}} Tr\left(F_{\mu\nu}\Phi_{54}F^{\mu\nu}\right).
\label{dim5}
\ea  
The details of this analysis will be presented elsewhere
\cite{mpr}. Suffice it to state that $|\eta_{1,2}| \sim {\cal{O}}(1)$ and
the interactions in eq.
(\ref{dim5}) lead to finite
corrections to the gauge couplings at the GUT-scale.  
The couplings of the left-right gauge group thus emerge from
one effective GUT-gauge coupling.  The
upshot of this is that with these additional contributions 
it is possible to lower $M_R$ to as low as
$10^9 - 10^{11}$ GeV as required in this model. The grand
unification scale is high:  $M_U \sim
10^{17-18}$ GeV and the model predicts a stable proton for all
practical purposes.

We expect that this model will have a natural extension to an $\bf
{E(6)}$-GUT wherein the matter multiplets and the singlet fields
will constitute the fundamental {\bf 27} representation of the
gauge group. 

In conclusion, we have presented a SUSY $\bf {SO(10)}$-based
model relying on a double see-saw mechanism which is (a)
consistent with the known neutrino masses and mixing, and (b) can
lead to a correct lepton asymmetry via the decays of sterile,
i.e.,  $\bf {SO(10)}$ singlet, neutrinos while remaining in
concordance with the gravitino constraint. The intermediate
scales are obtained through an RG analysis of the gauge coupling
running and are consistent with a long-lived proton. The model is
falsifiable through its prediction of doubly-charged Higgs bosons
within the reach of the LHC.

\vskip 5pt
{\bf Acknowledgements}: M.K.P. thanks the Harish-Chandra Research
Institute for hospitality.  The work is supported from funds of
the XIth Plan Neutrino project at HRI.  Computation facility of the HRI
Cluster project is also acknowledged.



\end{document}